\documentclass[11pt,a4paper,english,nofootinbib,,superscriptaddress]{revtex4}
\usepackage{lmodern}

\usepackage[T1]{fontenc}
\usepackage[latin9]{inputenc}
\usepackage{babel}
\usepackage{amsmath}
\usepackage{graphicx}
\usepackage{amssymb}
\usepackage{esint}
\usepackage[unicode=true, pdfusetitle,
 bookmarks=true,bookmarksnumbered=false,bookmarksopen=false,
 breaklinks=false,pdfborder={0 0 1},backref=false,colorlinks=false]
 {hyperref}
\usepackage{amsmath}
\usepackage{amssymb}
\def\b{\begin{equation}}
 \def\e{\end{equation}}

\makeatletter
\usepackage{latexsym}\usepackage{bm}

\makeatother

\begin{document}

\title{Inflation in Non-de Sitter Background with Coherent States}

\author{E. Yusofi}
\email{e.yusofi@iauamol.ac.ir}
\affiliation{Department of Physics, Ayatollah Amoli Branch, Islamic Azad University, Amol, Mazandaran, Iran}
\author{M. Mohsenzadeh}
\email{mohsenzadeh@qom-iau.ac.ir}
\affiliation{Department of Physics, Qom Branch, Islamic Azad University, Qom, Iran}
\author{M.R. Tanhayi}
\email{m_tanhayi@iauctb.ac.ir}
\affiliation{Department of Physics, Islamic Azad University, Central Tehran Branch, Tehran, Iran}

\date{\today}

\begin{abstract}

\noindent \hspace{0.35cm} We use the excited coherent states built over the initial non-de Sitter modes, to study the modification of spectra of primordial scalar fluctuation. Non-de Sitter modes are actually the asymptotic solution of the inflaton field equation[JHEP 09(2014) 020]. We build excited coherent states over the non-de Sitter modes and despite the lack of interactions in the Lagrangian, we find a non-zero one-point function. It is shown that the primordial non-Gaussianity resulting from excited-de Sitter modes depend both of time and background space-time. It is very tiny of order $(\leq{10}^{-24})$, at the Planck initial fixed time that confirmed by resent observations for single field inflation but it grows in the present epoch. Moreover, our results at the leading order are similar to what obtained with general initial states and in the dS limit leads to standard results [JCAP 1202(2012) 005]. We will show that the non-dS modes and its resulting spectrum are more usable for far past time limit.\\

\textbf{Pacs}: 04.62.+v, 04.25.-g, 98.80.Cq
\end{abstract}

\maketitle

\section{Introduction}

Inflation scenario was originally proposed in the 1980's in order to overcome some essential problems of the Big Bang theory \cite{inf1, inf2, inf3}. This scenario could potentially predict the scale invariance of the power spectrum and also the Gaussian distribution of primordial fluctuations. Remarkably, inflation is thought to be responsible both for the large-scale homogeneity of the universe and the small fluctuations that were the seeds for the formation of the large-scale structure (LSS). Actually,  very rapid expansion of the quantum fluctuations causes the inflated modes to be freezed in the super-horizon region so that they became as the classical fluctuations. These classical density fluctuations ${\Delta \rho}/{\rho}$ can appear in the form of observable temperature fluctuations i.e. ${\Delta T}/{T}$  in the cosmic microwave background (CMB)\cite{per4, per5, per6, per7, per8}.\\
The recent detections of observational cosmology such as scalar spectral index $n_{s}$ and the parameter of non-Gaussianity $f_{NL}$ of the CMB \cite{obs9, obs10} opens new windows to the physics of very early universe. Actually, the recent CMB results from Planck satellite and data from the Wilkinson Microwave Anisotropy Probe (WMAP)\cite{obs11}, impose an interesting constraint on the value of scalar spectral index approximately to be  $n_s = {0.9603} \pm {0.0073}$ at 95\% CL \cite{obs9}.  In \cite{man12}, we considered this constraint to show that the index of Hankel function $\nu$, as the general solution of inflaton field equation, lies in the range of $1.51\leq \nu \leq 1.53$ \footnote{Since the Bunch-Davies(BD) mode \cite{Bun33} is used just for pure de Sitter(dS) space-time i.e. $\nu = 3/2$, this range of $\nu$ motivates us to consider non-dS modes for quasi(excited)-dS space-time.} which it motivated us to the departure from BD mode to non-BD modes.\\
There has been a great deal of work focused on modifications of usual initial states and BD mode to calculate its effects on inflation, including $\alpha$-vacuum \cite{alf14}, general multi-mode squeezed states \cite{non15, non16, non17, non18, non19, non20}, homogeneous initial states \cite{Sin21}, general Gaussian and non-Gaussian initial states \cite{non22}, coherent states and $\alpha$-states \cite{San23, San24}, excited initial inflationary states \cite{Pab25}. Also, in \cite{San26, San27}, the effect of having thermal initial state on the power spectrum has already been considered. In \cite{Cal28, Agu29} a family of excited states that are indistinguishable from the BD mode at the level of two-point function or three-point function has been considered. Noting that, any non-linear effects in the expansion process or in the transformations between various stages of cosmic expansion process, will influence the final observable data \cite{non30, alf14}. In this paper we consider non-BD modes with non-linear part as the primary factors that could be an important source to generate scale-dependent power spectrum and non-Gaussianity. Such nontrivial modes have been used to calculate scale-dependent power spectrum with higher order of trans-Planckian corrections \cite{man12, Ash31}. On the other hand in \cite{man30, man31}, it is proved that by using these non-BD modes, one receives a renormalized theory of quantum field in which the symmetry of curved space-time is preserved. The main purpose of present work is to calculate the effects of non-dS modes with non-linear part on the spectra of fluctuations by employing of coherent states which is the generalization of \cite{San23}.\\
The rest of this paper proceed as follows: in Sec. 2 we review our recent work \cite{man12} which is about non-dS modes and motivations applying of it. In the section, first we study quantum fluctuations of scalar field during inflation in non-dS space-time and next we compute the power spectrum with BD mode and non-dS modes. In the main Sec. 3, by using non-dS modes, we generalize the results of \cite{San23} in coherent states by explicitly calculating one-point and two-point functions.  We also present some discussions about the primordial non-Gaussianity resulting from non-linear part of non-dS modes. Conclusions and outlook are given in the final section.

\section{Quantum Fluctuations of non-dS modes}

Let us start with the following action of minimally coupled real scalar field
\begin{equation}\label{act2}
S=\frac{1}{2}\int d^4x\sqrt{-g}\Big[{R}-(\nabla \phi)^2-2V(\phi)\Big],
\end{equation}
where $8\pi G=\hbar=1$ is used. The scalar perturbations in the inflaton field are described in terms of the gauge invariant comoving curvature perturbation ${R}$ which is given by \cite{per7, man12}
\begin{equation}\label{gau3}
{R}=\Psi+{\frac{H}{\dot{\bar{\phi}}}{\delta\phi}},
\end{equation}
where $\Psi$ is the spatial curvature perturbation. This form of scalar curvature in fact fixes the fictitious gauge modes \cite{per7}. Taking the dS flat metric and after expanding (\ref{act2}) to the second order one arrives to the Mukhanov action,
 \begin{equation}\label{act5}
S=\frac{1}{2}\int d^3xd{\tau} \Big[(\upsilon')^2+(\partial_i{\upsilon})^2-\frac{z''}{z}\upsilon^{2}\Big],
\end{equation} where the Mukhanov variable $\upsilon$ is defined as, $\upsilon=z{R}$ with $z^2=a^2\frac{\dot{\bar{\phi}}^2}{H^2}$, and the prime is the derivative with respect to conformal time $\tau$ and $a$ is the scale factor. Therefore the equation of motion in Fourier space is given by \cite{per7, Sin21},
\begin{equation} \label{muk7}  \upsilon''_{k}+(k^{2}-\frac{ z''}{z})\upsilon_{k}=0. \end{equation}
The quantization can indeed be done as follows
 \begin{equation}
 \label{mod8}
\upsilon\rightarrow \hat{\upsilon}=\int\frac{d\textbf{k}^3}{(2\pi)^3}\Big( \hat{a}_\textbf{k}\upsilon_k(\tau)e^{i\textbf{k}\cdot\textbf{x}}+ \hat{a}_\textbf{k}^\dagger\upsilon^{*}_k(\tau)e^{-i\textbf{k}\cdot\textbf{x}}\Big),
 \end{equation}
where $\hat{a}_\textbf{k}$ and $\hat{a}_\textbf{k}^\dagger$ are the annihilation and creation operators, respectively. Also, for the Fourier components $\upsilon_k $, we have following decomposition,
 \begin{equation}
 \label{mod9}
\upsilon_\textbf{k}\rightarrow \hat{\upsilon}_\textbf{k}=\frac{1}{\sqrt{2}}(\hat{a}_\textbf{k}\upsilon_k(\tau)+\hat{a}_{-\textbf{k}}^\dagger\upsilon^{*}_{-k}(\tau)).
 \end{equation}

\subsection{General Non-dS Inflationary Modes}

For the dynamical inflationary background, the equation (\ref{muk7}) change to general form as follows \cite{per7, man12, asl34},
\begin{equation} \label{Muk25}  \upsilon''_{k}+(k^{2}-\frac{2\alpha}{\tau^2})\upsilon_{k}=0, \end{equation}
where $\alpha$ is given by \cite{man12, asl34},
 \begin{equation}
 \label{alf26}  \alpha=\alpha(\nu)=\frac{4\nu^2-{1}}{8}.
 \end{equation}
  The general solutions of mode equation (\ref{Muk25}) can be written as \cite{per7, man12}: \begin{equation}
\label{Han22}
\upsilon_{k}=\frac{\sqrt{\pi \tau}}{2}\Big(A_{k}H_{\nu}^{(1)}(|k\tau|)+B_{k}H_{\nu}^{(2)}(|k\tau|)\Big), \end{equation} where
$ H_{\nu}^{(1, 2)} $ are the Hankel functions of the first and second kind, respectively.
Let us consider the general form of the mode function by expanding the Hankel functions up to the higher order of ${1}/{|k\tau|}$
 \begin{equation} \label{gen27}  \upsilon^{gen}_{k}(\tau, \nu)=A_{k}\frac{e^{-{i}k\tau}}{\sqrt{k}}\big(1-i\frac{\alpha}{k\tau}-\frac{\beta}{k^2\tau^2}-...\big)+
 B_{k}\frac{e^{{i}k\tau}}{\sqrt{k}}\big(1+i\frac{\alpha}{k\tau}-\frac{\beta}{k^2\tau^2}+...\big),
 \end{equation}
note that $\beta={\alpha(\alpha-1)}/{2}$. The positive frequency solutions of the mode equation (\ref{Muk25}) are given by \cite{man12}
\begin{equation}
\label{mod28}  \upsilon^{ND}_{k}=\frac{e^{-{i}k\tau}}{\sqrt{k}}\left(1-i\frac{\alpha}{k\tau}-\frac{\beta}{k^2\tau^2}-...\right).
 \end{equation}
These modes are non-linear in terms of both variables $\nu$ and $\tau$. If we consider these modes up to first order of $1/{k\tau}$, we will have
\begin{equation}
\label{mod29}  \upsilon^{ND1}_{k}=\frac{e^{-{i}k\tau}}{\sqrt{k}}\left(1-i\frac{\alpha}{k\tau}\right),
 \end{equation}
that the modes are linear in terms of $\tau$ and non-linear in terms of $\nu$ . In special case of the pure dS space-time($\nu={3}/{2}$), the general form of the mode functions (\ref{mod28}) leads to the exact BD mode:
  \begin{equation}
 \label{Bun29} \upsilon_{k}^{BD}=\frac{1}{\sqrt{k}}(1-\frac{i}{k\tau})e^{-ik\tau}.
\end{equation}
For this case, one has $a(t)=e^{Ht}$, or $ a(\tau)=-\frac{1}{{H}\tau}$, with $H=constant$ for very early universe. In \cite{man12}, an asymptotically flat excited solution (\ref{mod28}) has been considered during the inflation in which the best values of $\nu$ which are confirmed with the latest observational data \cite{obs9}, is $1.51\leq \nu \leq 1.53$. This result motivated us to the departure from BD mode to non-BD modes. Therefor we use non-dS modes (\ref{mod28}) instead dS mode as the fundamental modes for our calculation in the next sections.

\subsection{Scale-Dependent Power Spectrum with non-dS Modes}

As the mentioned in above, the observations of CMB and LSS tell us conclusively that the cosmic inflation is described by nearly dS space-time and the power spectrum of the fluctuations produced during inflation is nearly scale invariant (i.e. $n_{s}\approx1$) \cite{obs9, obs10}. Motivated by this fact, we use the non-dS modes instead of the usual BD mode. Using the modes introduced in (\ref{mod28}) one obtains
\begin{equation}
 \label{pow31} P_{{R}}=\frac{1}{2a^2}(\frac{H^2}{\dot{\bar{\phi}}^2})|\upsilon_{k}^{ND}(\tau)|^{2}.
\end{equation}
For the super-horizon limit $k\tau\ll1$, the following modified power spectrum in terms of $\nu$ has been calculated \cite{man12}
\begin{equation}
\label{del32}  \Delta_{{R}}^{2}=\frac{H^2}{(2\pi)^{2}}(\frac{H^2}{\dot{\bar{\phi}}^2})\left[\frac{1}{2}(\frac{2\nu+1}{2\nu-1})+(2\nu+1)^{2} \frac{(4\nu^{2}-9)^{2}}{64k^{2}\tau^{2}}+...\right].
\end{equation}
It is worth to mention that the following relation for $\tau$ is used \cite{asl34},
\begin{equation}
 \label{eta33} \tau=\frac{-1}{aH}\left(\nu-\frac{1}{2}\right),
\end{equation}
where, we have assumed that $\nu\approx{3/2}+\epsilon$ with a constant slow-roll parameter $ \epsilon $ \cite{asl34}. This indicates that the conformal time $\tau$ can be depends on $\nu$. On the other hand, for ND1 modes (\ref{mod29}) one obtains the modified power spectrum as,
  \begin{equation} \label{del362}
 \Delta_{{R}}^{2}=\frac{H^2}{(2\pi)^{2}}(\frac{H^2}{\dot{\bar{\phi}}^2})\left[\frac{1}{2}(\frac{2\nu+1}{2\nu-1})\right].
\end{equation}
Where the curvature perturbation is given by ${R}_{k}(\tau)=\frac{\upsilon_{k}(\tau)}{a}({H}/{{\dot{\bar{\phi}}}})$. Also, $P_{{R}}$ and $\Delta_{{R}}^{2}$ are the power spectrum and dimensionless power spectrum \cite{man12}, respectively.\\
For pure dS phase namely $\nu={3}/{2}$, one obtains
    \begin{equation}
 \label{alf Bunch mode3/2} \alpha=1,\quad  \epsilon=\frac{-\dot{H}}{H^{2}}=0,
\end{equation}
 which is the same as the result of using of the BD mode and (\ref{del32}) reduces to the standard scale invariant power spectrum. Therefore, with regarding the above results, it is deduced the utilizing of non-dS modes, can be consider as the primary factor for generating the scale-dependent power spectrum and non-Gaussian effects in CMB. Similar to our results, the non-linear corrections of power spectrum obtained from previous conventional methods [35- 44]. We are going to investigate this issue by making use of coherent state in the next section.

\section{Calculation of Spectra in Coherent States}

 Since, we do not know anything about the physical states before inflation, any excited state is as good an initial state as the vacuum state.  Excited states can be made by using creation operators $\hat{a}_{k}^\dag$ over the vacuum state $|0\rangle$,
\begin{equation} \label{coh1}
 |\psi\rangle=\frac{1}{\sqrt{n1!n2!...}}[(\hat{a}_{k1}^\dag)^{n1}(\hat{a}_{k2}^\dag)^{n2}...]|0\rangle,
\end{equation}
 where for the excited coherent state, we have $|\psi\rangle\equiv|C\rangle$ and the coherent state $|C\rangle$ is defined as
 \begin{equation}
 \hat{a}_{k}|C\rangle=C(\textbf{k})|C\rangle, \quad  and \quad  \langle {C}|\hat{a}_{k}^\dag=\langle {C}|C^{*}(\textbf{k}).
\end{equation}
In fact, the coherent state are the quantum states that well describe the quantum harmonic oscillator whose dynamics resembles the classical harmonic oscillator behaviors.\\
If we build coherent state over the BD mode $|0\rangle$ in (\ref{coh1}), the homogeneity in the large-scale as a physical constrain leads to deduce the one-point function of $\hat{R}$ in the super-horizon limit $|k\tau|\ll{1}$ to be zero\footnote{This condition only for the BD mode $|0\rangle$ is true, but for a general initial state and for interacting quantum field, the operator $\hat{R}_\textbf{k}$ can have a non-vanishing expectation value \cite{San24}.},
\begin{equation} \label{con53}
 \langle{C}|\hat{R}_\textbf{k}(\tau)|{C}\rangle=0,
\end{equation}
where $\hat{R}_{\textbf{k}}$ is defined by,
\begin{equation}
 \hat{R}_{\textbf{k}(\tau)}=\frac{1}{\sqrt{2}}[\hat{a}_\textbf{k}{R}^{*}_{k}(\tau)+\hat{a}_{-\textbf{k}}^\dagger {R}_{k}(\tau)],
\end{equation}
where,
\begin{equation}
{R}^{BD}_{k}(\tau)=(\frac{H}{a\dot{\bar{\phi}}})\frac{1}{\sqrt{k}}(1-\frac{i}{k\tau})e^{-ik\tau}.
\end{equation}
So, the constrain (\ref{con53}) leads to the following condition
\begin{equation} \label{con1}
C^{*}(-\textbf{k})=C(\textbf{k}),
\end{equation}
with these coherent states as the initial excited states, we can compute two-point function as follows \cite{San23},
\begin{equation} \label{pow511}
\langle\hat{R}_\textbf{k}(\tau)\hat{R}_{\textbf{k}'}(\tau)\rangle=\langle{C}|\hat{R}_\textbf{k}(\tau)\hat{R}_{\textbf{k}'}(\tau)|{C}\rangle=\frac{1}{2}(2\pi)^{3}\frac{H^4}{\dot{\bar{\phi}}^{2}k^3}\delta^{3}(k+k')
\end{equation}
Note that the calculated power spectrum (\ref{pow511}) with coherent state $|C\rangle$, is exactly like to the calculated power spectrum with BD mode $|0\rangle$. Therefore one has,
\begin{equation}
\langle{C}|\hat{R}_\textbf{k}(\tau)\hat{R}_{\textbf{k}'}(\tau)|{C}\rangle=\langle{0}|\hat{R}_\textbf{k}(\tau)\hat{R}_{\textbf{k}'}(\tau)|{0}\rangle.
\end{equation}
\section{Coherent states over the non-dS Modes}
The present observations of the CMB temperature inhomogeneities indicates the presence of almost scale-invariant spectrum of curvature perturbations \cite{obs9}. On the other hand theoretically, temperature fluctuations of CMB and LSS are directly originated from the curvature perturbations produced during inflation.\\ \emph{The correction terms in non-dS modes was very tiny (nonzero) in the early time but can be grow in the later time. As it is known at the early time in short distance regime the energy of the universe was very high and the potential of the universe has located in the state with maximal symmetry, therefore these tiny corrections terms of initial modes at early time limit may play a role as initial sources to spontaneous symmetry bricking, bubble nucleation, and creation of inflating universe. In the context of effective field theory, in short distance scales and in the sub-horizon limit the initial symmetry may be broken by such non-linear effects to outburst and propagation quantum fields and particles \cite{man46} in the super-horizon scale to formation of large galaxies and galaxies cluster. To verify this claim, let us first build coherent excited states over non-dS modes and examine the effects of the correction terms in the sub-horizon and super-horizon limit.}\\
 It is shown in the previous section that the one-point function for coherent states built over BD mode(special case of non-dS modes with $\nu=\frac{3}{2}$ and linear order of $\frac{1}{k\tau}$ )  in the super-horizon limit is zero under the constraint (\ref{con53}). In this subsection we build coherent excited states over non-dS modes and we want to compute one-point and two-point functions, respectively. Note that all coefficients of $\frac{1}{(k\tau)^{n}}$, $n\geq{2}$ in (\ref{mod28}) are important but for simplification of calculations and results, we stop the expansion up to the second order.
\subsection{Calculation of One-Point Function}
  We can write
\begin{equation} \label{sahm1}
{\langle{C}|\hat{R}_{k}(\tau)|{C}\rangle}_{ND}={\langle{C}|\hat{R}_{k}(\tau)|{C}\rangle}_{L}+{\langle{C}|\hat{R}_{k}(\tau)|{C}\rangle}_{NL}.
\end{equation}
Where the first term on the right hand side corresponds to the linear part of non-dS modes and the second term corresponds to the contribution of non-linear part. If we apply condition (\ref{con1}), the first term of the above equation will be zero, however the second term is non-zero leading to
\begin{equation} \label{one11}
{\langle{C}|\hat{R}_{k}(\tau)|{C}\rangle}_{ND}={\langle{C}|\hat{R}_{k}(\tau)|{C}\rangle}_{NL}=\frac{-2C(\textbf{k})\beta}{k^{2}\tau^{2}}+... .
\end{equation}In other words the second or higher order terms of $1/k{\tau}$ inserted in the non-dS modes act like general initial state or interacting field in effective field theory method. We would like to note that in the presence of three-point interaction \cite{mal45}, one has $ \langle{C}|\hat{R}_{k}(\tau\rightarrow0)|{C}\rangle\neq0$ \cite{San23, San24}, while in the present method, just because of the presence of higher order corrections terms of $1/k\tau$, we can obtain non-zero one-point function. This can leads to a non-zero three-point function as the following,
 \begin{equation} \label{one13}
  \langle\hat{R}^{NL}_\textbf{k1}(\tau)\hat{R}^{NL}_\textbf{k2}(\tau)\hat{R}^{NL}_\textbf{k3}(\tau)\rangle\neq{0}.
  \end{equation} Actually, the non-dS modes can play the role of redefined fields \cite{San23, San24} in the interaction picture. In the next subsections similar to the approach of \cite{San23}, we will compute power spectrum with non-dS modes.

\subsection{Calculation of Two-Point Function}

The power spectrum is calculated as follows
\begin{equation}
\langle{C}|\hat{R}_\textbf{k}(\tau)\hat{R}_{\textbf{k}'}(\tau)|{C}\rangle=
(\frac{H}{a\dot{\bar{\phi}}})^{2}\langle{C}|\hat{\upsilon}_\textbf{k}(\tau)\hat{\upsilon}_{\textbf{k}'}(\tau)|{C}\rangle
\end{equation}
with the non-dS modes of (\ref{mod28}), one obtains
$$
\langle{C}|\hat{\upsilon}^{ND}_\textbf{k}(\tau)\hat{\upsilon}^{ND}_{\textbf{k}'}(\tau)|{C}\rangle=\frac{1}{2}(2\pi)^{3}\delta^{3}(k+k')|\upsilon^{ND}_{k}(\tau)|^{2}
+A(\textbf{k},\textbf{k}'){\upsilon^{ND}}^{*}_{k}{\upsilon^{ND}}^{*}_{k'}+A(\textbf{-k},\textbf{-k}')\upsilon^{ND}_{k}\upsilon^{ND}_{k'}$$
\begin{equation}
+B(\textbf{-k},\textbf{k}')\upsilon^{ND}_{k}{\upsilon^{ND}}^{*}_{k'}
+B(\textbf{k},\textbf{-k}'){\upsilon^{ND}}^{*}_{k}\upsilon^{ND}_{k'},
\end{equation}
where we can write,
$$
A(\textbf{k},\textbf{k}')=\frac{1}{2}\langle{C}|\hat{a}_{\textbf{k}}\hat{a}_{\textbf{k}'}|{C}\rangle \quad, \quad A(\textbf{-k},\textbf{-k}')=\frac{1}{2}\langle{C}|\hat{a}^{\dag}_{\textbf{-k}}\hat{a}^{\dag}_{\textbf{-k}'}|{C}\rangle $$
\begin{equation}
B(\textbf{k},\textbf{-k}')=\frac{1}{2}\langle{C}|\hat{a}_{\textbf{k}}\hat{a}^{\dag}_{\textbf{-k}'}|{C}\rangle \quad, \quad B(\textbf{-k},\textbf{k}')=\frac{1}{2}\langle{C}|\hat{a}^{\dag}_{\textbf{-k}}\hat{a}_{\textbf{k}'}|{C}\rangle.
\end{equation}By introducing $ k_{*}=\sqrt{kk'}, \bar{k}=k+k'$ , $\delta{k}=k-k'$ and after making use of the non-dS modes (\ref{mod28}), the two-point function is found to be
$$
\langle{C}|\hat{\upsilon}^{ND}_\textbf{k}(\tau)\hat{\upsilon}^{ND}_{\textbf{k}'}(\tau)|{C}\rangle={\langle{C}|\hat{\upsilon}_\textbf{k}(\tau)\hat{\upsilon}_{\textbf{k}'}(\tau)|{C}\rangle}_{ND}=\frac{1}{2}(2\pi)^{3}\delta^{3}(k+k')\frac{1}{k}( 1+\frac{\alpha}{k^2\tau^2}+\frac{\beta^2}{k^4\tau^4}+...)$$
$$+A(\textbf{k},\textbf{k}')\frac{e^{i\bar{k}{\tau}}}{k_{*}}\left[1+\frac{i\alpha}{\tau}(\frac{\bar{k}}{k_{*}^{2}})-\frac{\alpha^{2}}{k_{*}^{2}\tau^{2}}
-\frac{\beta}{\tau^{2}}(\frac{\bar{k}^{2}-2k_{*}^{2}}{k_{*}^{4}})-\frac{i\alpha\beta}{k_{*}^{2}\tau^{3}}(\frac{\bar{k}}{k_{*}^{2}})+\frac{\beta^{2}}{k_{*}^{4}\tau^{4}}+...\right]
$$
$$+B(\textbf{-k},\textbf{k}')\frac{e^{-i\delta{k}{\tau}}}{k_{*}}\left[1+\frac{i\alpha}{\tau}(\frac{\delta{k}}{k_{*}^{2}})+\frac{\alpha^{2}}{k_{*}^{2}\tau^{2}}
-\frac{\beta}{\tau^{2}}(\frac{\bar{k}^{2}-2k_{*}^{2}}{k_{*}^{4}})+\frac{i\alpha\beta}{k_{*}^{2}\tau^{3}}(\frac{\delta{k}}{k_{*}^{2}})+\frac{\beta^{2}}{k_{*}^{4}\tau^{4}}+...\right] $$
$$
+B(\textbf{k},\textbf{-k}')\frac{e^{i\delta{k}{\tau}}}{k_{*}}\left[1-\frac{i\alpha}{\tau}(\frac{\delta{k}}{k_{*}^{2}})+\frac{\alpha^{2}}{k_{*}^{2}\tau^{2}}
-\frac{\beta}{\tau^{2}}(\frac{\bar{k}^{2}-2k_{*}^{2}}{k_{*}^{4}})-\frac{i\alpha\beta}{k_{*}^{2}\tau^{3}}(\frac{\delta{k}}{k_{*}^{2}})+\frac{\beta^{2}}{k_{*}^{4}\tau^{4}}+...\right]
$$
\begin{equation}
\label{gen48} +A(\textbf{-k},\textbf{-k}')\frac{e^{-i\bar{k}{\tau}}}{k_{*}}\left[1-\frac{i\alpha}{\tau}(\frac{\bar{k}}{k_{*}^{2}})-\frac{\alpha^{2}}{k_{*}^{2}\tau^{2}}
-\frac{\beta}{\tau^{2}}(\frac{\bar{k}^{2}-2k_{*}^{2}}{k_{*}^{4}})+\frac{i\alpha\beta}{k_{*}^{2}\tau^{3}}(\frac{\bar{k}}{k_{*}^{2}})+\frac{\beta^{2}}{k_{*}^{4}\tau^{4}}+...\right], \end{equation}Equivalently, shares arising from the linear and non-linear terms of $1/k\tau$ of non-dS modes, similar to (\ref{sahm1}), the above relation can be written as
\begin{equation}\label{two49}
{\langle{C}|\hat{\upsilon}_\textbf{k}(\tau)\hat{\upsilon}_{\textbf{k}'}(\tau)|{C}\rangle}_{ND}=
{\langle{C}|\hat{\upsilon}_\textbf{k}(\tau)\hat{\upsilon}_{\textbf{k}'}(\tau)|{C}\rangle}_{L}+{\langle{C}|\hat{\upsilon}_\textbf{k}(\tau)\hat{\upsilon}_{\textbf{k}'}(\tau)|{C}\rangle}_{NL}.
\end{equation}Where for the linear part of non-dS modes (\ref{mod28}), one has
$$
{\langle{C}|\hat{\upsilon}_\textbf{k}(\tau)\hat{\upsilon}_{\textbf{k}'}(\tau)|{C}\rangle}_{L}=\frac{1}{2}(2\pi)^{3}\delta^{3}(k+k')\frac{1}{k}( 1+\frac{\alpha}{k^2\tau^2})$$
$$+A(\textbf{k},\textbf{k}')\frac{e^{i\bar{k}{\tau}}}{k_{*}}\left[1+\frac{i\alpha}{\tau}(\frac{\bar{k}}{k_{*}^{2}})-\frac{\alpha^{2}}{k_{*}^{2}\tau^{2}}
\right]+B(\textbf{-k},\textbf{k}')\frac{e^{-i\delta{k}{\tau}}}{k_{*}}\left[1+\frac{i\alpha}{\tau}(\frac{\delta{k}}{k_{*}^{2}})+\frac{\alpha^{2}}{k_{*}^{2}\tau^{2}}
\right] $$
\begin{equation}\label{two50}
+B(\textbf{k},\textbf{-k}')\frac{e^{i\delta{k}{\tau}}}{k_{*}}\left[1-\frac{i\alpha}{\tau}(\frac{\delta{k}}{k_{*}^{2}})+\frac{\alpha^{2}}{k_{*}^{2}\tau^{2}}
\right]+A(\textbf{-k},\textbf{-k}')\frac{e^{-i\bar{k}{\tau}}}{k_{*}}\left[1-\frac{i\alpha}{\tau}(\frac{\bar{k}}{k_{*}^{2}})-\frac{\alpha^{2}}{k_{*}^{2}\tau^{2}}
\right],
\end{equation} and the non-linear part of modes (\ref{mod28}), is as follows
 $$
{\langle{C}|\hat{\upsilon}_\textbf{k}(\tau)\hat{\upsilon}_{\textbf{k}'}(\tau)|{C}\rangle}_{NL}=\frac{1}{2}(2\pi)^{3}\delta^{3}(k+k')\frac{1}{k}( \frac{\beta^2}{k^4\tau^4})$$
$$+A(\textbf{k},\textbf{k}')\frac{e^{i\bar{k}{\tau}}}{k_{*}}\left[-\frac{\beta}{\tau^{2}}(\frac{\bar{k}^{2}-2k_{*}^{2}}{k_{*}^{4}})-\frac{i\alpha\beta}{k_{*}^{2}\tau^{3}}(\frac{\bar{k}}{k_{*}^{2}})+\frac{\beta^{2}}{k_{*}^{4}\tau^{4}}
\right]$$
$$+B(\textbf{-k},\textbf{k}')\frac{e^{-i\delta{k}{\tau}}}{k_{*}}\left[-\frac{\beta}{\tau^{2}}(\frac{\bar{k}^{2}-2k_{*}^{2}}{k_{*}^{4}})+\frac{i\alpha\beta}{k_{*}^{2}\tau^{3}}(\frac{\delta{k}}{k_{*}^{2}})+\frac{\beta^{2}}{k_{*}^{4}\tau^{4}}
\right] $$
$$ +B(\textbf{k},\textbf{-k}')\frac{e^{i\delta{k}{\tau}}}{k_{*}}\left[-\frac{\beta}{\tau^{2}}(\frac{\bar{k}^{2}-2k_{*}^{2}}{k_{*}^{4}})-\frac{i\alpha\beta}{k_{*}^{2}\tau^{3}}(\frac{\delta{k}}{k_{*}^{2}})+\frac{\beta^{2}}{k_{*}^{4}\tau^{4}}
\right]$$
\begin{equation}\label{two51}
+A(\textbf{-k},\textbf{-k}')\frac{e^{-i\bar{k}{\tau}}}{k_{*}}\left[-\frac{\beta}{\tau^{2}}(\frac{\bar{k}^{2}-2k_{*}^{2}}{k_{*}^{4}})+\frac{i\alpha\beta}{k_{*}^{2}\tau^{3}}(\frac{\bar{k}}{k_{*}^{2}})+\frac{\beta^{2}}{k_{*}^{4}\tau^{4}}
\right],
\end{equation}If we consider special case $\nu=3/2$  or  $\alpha=1$, we will have $\beta={0}$, and ${\langle{C}|\hat{\upsilon}_\textbf{k}(\tau)\hat{\upsilon}_{\textbf{k}'}(\tau)|{C}\rangle}_{NL}={0}$, so we obtain,
$$
{\langle{C}|\hat{\upsilon}_\textbf{k}(\tau)\hat{\upsilon}_{\textbf{k}'}(\tau)|{C}\rangle}_{L}=\frac{1}{2}(2\pi)^{3}\delta^{3}(k+k')\frac{1}{k}( 1+\frac{1}{k^2\tau^2})$$
$$+A(\textbf{k},\textbf{k}')\frac{e^{i\bar{k}{\tau}}}{k_{*}}\left[1+\frac{i}{\tau}(\frac{\bar{k}}{k_{*}^{2}})-\frac{1}{k_{*}^{2}\tau^{2}}
\right]+B(\textbf{-k},\textbf{k}')\frac{e^{-i\delta{k}{\tau}}}{k_{*}}\left[1+\frac{i}{\tau}(\frac{\delta{k}}{k_{*}^{2}})+\frac{1}{k_{*}^{2}\tau^{2}}
\right] $$
\begin{equation}\label{lin52}
+B(\textbf{k},\textbf{-k}')\frac{e^{i\delta{k}{\tau}}}{k_{*}}\left[1-\frac{i}{\tau}(\frac{\delta{k}}{k_{*}^{2}})+\frac{1}{k_{*}^{2}\tau^{2}}
\right]+A(\textbf{-k},\textbf{-k}')\frac{e^{-i\bar{k}{\tau}}}{k_{*}}\left[1-\frac{i}{\tau}(\frac{\bar{k}}{k_{*}^{2}})-\frac{1}{k_{*}^{2}\tau^{2}}
\right],\end{equation}Noting that this result is the same as what obtained in \cite{San23} where the general initial states with BD mode has been used whereas we used the coherent states with non-dS modes. For the super-horizon limit, by applying condition (\ref{con1}), one  achieves following constrain,

\begin{equation}\label{con521}
A(\textbf{k},\textbf{k}')= A(\textbf{-k},\textbf{-k}')=B(\textbf{k},\textbf{-k}')=B(\textbf{-k},\textbf{k}')=\frac{1}{2}C(\textbf{k})C({\textbf{k}}').
\end{equation}Note that by consideration above condition, the sum of the last four terms of the equation (\ref{lin52}) is equal to zero and we obtain
\begin{equation}\label{gen54}
{\langle{C}|\hat{\upsilon}_\textbf{k}(\tau)\hat{\upsilon}_{\textbf{k}'}(\tau)|{C}\rangle}_{L}\approx\frac{1}{2}(2\pi)^{3}\delta^{3}(k+k') \frac{1}{k^3\tau^2}.
\end{equation}Consequently, one receives scale invariant power spectrum. Considering the general case of ${\nu}> 3/2$ or  $\alpha\neq1$, leads to $\beta\neq{0}$. In this case, one should consider the contribution of non-linear terms of $1/k\tau$ in the calculation of the power spectrum. As a result, the non-zero contribution of non-linear corrections for power spectrum is obtained as, ${\langle{C}|\hat{\upsilon}_\textbf{k}(\tau)\hat{\upsilon}_{\textbf{k}'}(\tau)|{C}\rangle}_{NL}\neq{0}$. Therefore, by applying condition (\ref{con521}) for the super-horizon limit, one obtains
$${\langle{C}|\hat{\upsilon}_\textbf{k}(\tau)\hat{\upsilon}_{\textbf{k}'}(\tau)|{C}\rangle}_{NL}\approx\frac{1}{2}(2\pi)^{3}\delta^{3}(k+k')\frac{1}{k}( \frac{\beta^2}{k^4\tau^4}+...)$$
\begin{equation}\label{non55}
+2C(\textbf{k})C({\textbf{k}}')\left[-\frac{\beta}{\tau^{2}}(\frac{\bar{k}^{2}-2k_{*}^{2}}{k_{*}^{4}})+\frac{\beta^{2}}{k_{*}^{4}\tau^{4}}+...
\right].
\end{equation}Finally, we obtain two-point function for general non-dS modes in the super-horizon limit as,
$${\langle{C}|\hat{\upsilon}_\textbf{k}(\tau)\hat{\upsilon}_{\textbf{k}'}(\tau)|{C}\rangle}_{ND}\approx\frac{1}{2}(2\pi)^{3}\delta^{3}(k+k')\frac{1}{k}(\frac{\alpha}{k^2\tau^2}+ \frac{\beta^2}{k^4\tau^4}+...)$$
\begin{equation}\label{non55}
+2C(\textbf{k})C({\textbf{k}}')\left[-\frac{\beta}{\tau^{2}}(\frac{\bar{k}^{2}-2k_{*}^{2}}{k_{*}^{4}})+\frac{\beta^{2}}{k_{*}^{4}\tau^{4}}+...
\right].
\end{equation}
Note that, resulting from non-linear corrections terms of non-dS modes and according to the results obtained for the one-point function(\ref{one11})it looks for two-point function (\ref{non55}) and for the three-point function (\ref{one13}), the corrections can be of order $(\leq{1/{(k\tau)^{4}}})$ and $(\leq{1/{(k\tau)^{6}}}$, respectively. For initial fixed time $ \tau_{0}=-\frac{M_{Pl}}{Hk} $, if $M_{Pl}$ is the Planck scale,$\frac{H}{M_{Pl}}$ is at most $10^{-4}$\cite{tra44}, and we can obtain $(\leq{1/{(k\tau_{0})^{4}}})\approx(\frac{H}{M_{Pl}})^{4}\approx{10}^{-16}$ and $(\leq{1/{(k\tau_{0})^{6}}}\approx(\frac{H}{M_{Pl}})^{6}\approx{10}^{-24}$, that indicates the resulting corrections are very tiny at Planck time $\tau_{0}$, and it imply the nearly scale-invariant power spectrum and almost Gaussian distribution in CMB\footnote{ An adiabatic, Gaussian and nearly scale-independent scalar power spectrum has also been confirmed by Wilkinson Microwaves Anisotropy Probe 9-year data  \cite{map47} and Planck data released in 2013 \cite{obs9}.}. This final result emphasize that our non-dS modes are more usable for far past time limit.
\section{Conclusions}

It is known that any deviation of Bunch-Davies initial state in inflation would lead to corrections of the power spectrum. In this paper we have employed the initial non-dS modes to study the corrections of the spectra mainly coming from the comoving curvature perturbation. Non-dS modes are actually the asymptotic expansion of the Hankel function with index $\nu\approx{3/2+\epsilon}$, as the general solution of inflaton field equation at very early universe. This followed from the fact that the space-time of inflation are indeed nearly dS space-time but not exact dS space-time, this motivates one to deviate dS mode and consider general non-dS modes.  As a matter of fact when one carries out the renormalization with the non-dS modes the symmetry of the space-time is preserved as long as the dS space-time is supposed to be as the background \cite{man30}.  Explicitly calculations showed that using such modes leads to tiny and non-zero one-point function for far past time. This result may mean that the anisotropy of CMB radiation, can be originated from the non-linearity of the initial states. Note that the corrections obtained with non-dS modes very closely related to the non-linear corrections that have calculated with general initial states in ref.\cite{San23}, and in fact, be complementary corrections were obtained with linear BD mode. It looks the non-linear part of non-dS modes, can play the role of redefined fields in the interaction picture to calculate of three-point functions. Finally, it is shown that the primordial non-Gaussianity in single field inflation resulting from initial non-dS states at the Planck scale, might be very tiny that confirmed by resent observation. Moreover, our results at the leading order were similar to what obtained with general initial states and in the dS limit leaded to standard result.\\ Although, we examine both of sub-horizon and super-horizon limits, but we emphasize that our non-trivial modes and resulting spectra are more reasonable and suitable for far past time limit, $k\tau\gg1$, specially for initial fixed time $\tau_0$. In future work we will study what interaction terms and what form of potential in the Lagrangian can lead to the corrections terms in the the non-dS mods.\\

\noindent{\bf{Acknowlegements}}: We would like to thank Mohammad Vahid Takook, Sandipan Kundu, Hamed Pejhan for constructive conversations and comments. This work has been supported by the Islamic Azad University, Ayatollah Amoli Branch, Amol, Mazandaran, Iran.


\begin{thebibliography}{a}
\bibitem{inf1} A. H. Guth, \emph{The Inflationary Universe: A Possible Solution to the Horizon and Flatness Problems}, Phys. Rev. D 23, 347 (1981).

\bibitem{inf2} A. Linde, \emph{Particles Physics and Inflationary Cosmology}, Harwood Academic, Reading (1991).

\bibitem{inf3} A. R. Liddle, \emph{An Introduction to Cosmological Inflation}, (1999), [astro-ph/9901124v1].

\bibitem{per4} A. A. Starobinsky, \emph{Dynamics of Phase Transition in The New Inflationary Universe Scenario and Generation of Perturbations}, Phys. Lett. B 117, 175 (1982).

\bibitem{per5} S. W. Hawking, \emph{The Development of Irregularities in a Single Bubble Inflationary Universe}, Phys. Lett. B 115, 295 (1982).

\bibitem{per6} A. H. Guth and S. Y. Pi, \emph{Fluctuations in the New Inflationary Universe}, Phys. Rev. Lett. 49, 1110 (1982).

\bibitem{per7} D. Baumann, \emph{TASI Lectures on Inflation}, TASI 2009, [hep-th/0907.5424].

\bibitem{per8} V. Mukhanov, \emph{Physical Foundations of Cosmology}, Mar., 2001.
\bibitem{obs9} P. A. R. Ade, et. al., \emph{Planck 2013 Results. XXII. Constraints on Inflation},[astro-ph/1303.5082]; \\S. Unnikrishnana and V. Sahni, \emph{Resurrecting Power Law Inflation in the Light Of Planck Results}, JCAP 10 (2013) 063,[astro-ph/1305.5260].
\bibitem{obs10} P. A. R. Ade, et. al., \emph{Planck 2013 Results. XXIV. Constraints on primordial non-Gaussianity},[astro-ph/1303.5084].
\bibitem{obs11} E. Komatsu et al., \emph{Five-Year Wilkinson Microwave Anisotropy Probe (WMAP) Observations: Cosmological Interpretation}, Astrophys. J. Suppl.,
 [astro-ph/0803.0547].
\bibitem{man12} E. Yusofi and M. Mohsenzadeh, \emph{Scale-dependent power spectrum from initial excited-de Sitter modes}, JHEP09 (2014) 020,[astro-ph/1402.6968], E. Yusofi and M. Mohsenzadeh, \emph{An Asymptotic Method for Selection of Inflationary Modes}, Mod. Phys. Lett. A 30, 9 (2015) 1550041.
\bibitem{alf14} W. Xue and B. Chen, \emph{$\alpha$-Vacuum and Inflationary Bispectrum}, Phys. Rev. D 79(2009), [hep-th/0806.4109].
\bibitem{non15} F. Nitti, M. Porrati, and J.-W. \emph{Rombouts,Naturalness in cosmological initial conditions}, Phys.Rev. D72 (2005) 063503, [hep-th/0503247].

\bibitem{non16} M. Porrati, \emph{Effective field theory approach to cosmological initial conditions: Self-consistency
bounds and non-Gaussianities}, [hep-th/0409210].

\bibitem{non17} R. Holman and A. J. Tolley, \emph{Enhanced Non-Gaussianity from Excited Initial States}, JCAP 0805 (2008) 001, [hep-th/0710.1302].

\bibitem{non18} P. D. Meerburg, J. P. van der Schaar, and P. S. Corasaniti, \emph{Signatures of Initial State Modifications on Bispectrum Statistics}, JCAP 0905 (2009) 018, [hep-th/0901.4044].

\bibitem{non19} J. Ganc, \emph{Calculating the local-type $f_{NL}$ for slow-roll inflation with a non-vacuum initial state}, Phys.Rev. D84 (2011) 063514, [astro-ph/1104.0244].

\bibitem{non20} I. Agullo and S. Shandera, \emph{Large non-Gaussian Halo Bias from Single Field Inflation}, JCAP 1209 (2012) 007, [astro-ph/1204.4409].
\bibitem{Sin21} S. Bahrami, E. E. Flanagan, \emph{Primordial non-Gaussianities in single field inflationary models with nontrivial initial states},[astro-ph/1310.4482].
\bibitem{non22} N. Agarwal, R. Holman, A. J. Tolley, and J. Lin, \emph{Effective field theory and non-Gaussianity from general inflationary states}, JHEP 1305 (2013) 085, [hep-th/1212.1172], C.P. Burgess, James M. Cline, F. Lemieux, R. Holman, \emph{Are inflationary predictions sensitive to very high-energy physics?}, JHEP 0302 (2003) 048, [hep-th/0210233].

\bibitem{San23} S. Kundu,\emph{Inflation with General Initial Conditions for Scalar Perturbations}, JCAP 1202 (2012) 005, [astro-ph/1110.4688].

\bibitem{San24} S. Kundu, \emph{Non-Gaussianity Consistency Relations, Initial States and Back-reaction}, [astro-ph/1311.1575].
\bibitem{Pab25} A. Aravind, D. Lorshbough, and S. Paban, \emph{Non-Gaussianity from Excited Initial Inflationary States}, JHEP 07(2013) 076, [hep-th/1303.1440].
\bibitem{San26} K. Bhattacharya, S. Mohanty and R. Rangarajan, \emph{Temperature of the inflaton and duration of inflation from WMAP data}, Phys. Rev. Lett. 96, 121302 (2006). [hep-ph/0508070].
\bibitem{San27} P. Ferreira, J. Magueijo, \emph{Observing the temperature of the Big Bang through large-scale structure}, Phys. Rev. D78, 061301 (2008), [astro-ph/0708.0429].
\bibitem{Cal28} A. Ashoorioon and G. Shiu, \emph{A Note on Calm Excited States of Inflation}, JCAP 1103 (2011)025, [arXiv:1012.3392].
\bibitem{Agu29} I. Agullo and L. Parker,\emph{ Non-gaussianities and the Stimulated creation of quanta in the inflationary universe}, Phys.Rev. D83 (2011) 063526, [astro-ph/1010.5766].
\bibitem{non30} Y. Takamizu and T. Kobayashi, \emph{Nonlinear superhorizon curvature perturbation in generic single-field inflation}, Prog. Theor. Exp. Phys. (2013) 2013 (6): 063E03 doi:10.1093/ptep/ptt033 First published online June 1, 2013 (17 pages).
\bibitem{Ash31} A. Ashoorioon et al., \emph{Running of the Spectral Index and Violation of the Consistency Relation Between Tensor and Scalar Spectra from trans-Planckian Physics}, Nucl.Phys. B727 (2005) 63-76, [gr-qc/0504135].
\bibitem{man30} M. Mohsenzadeh, M. R. Tanhayi and E. Yusofi, \emph{Power spectrum with auxiliary fields in de Sitter space}, Eur. Phys. J. C (2014) 74:2920, DOI 10.1140/epjc/s10052-014-2920-5, [hep-th/1306.6722].
\bibitem{man31} E. Yusofi and M. Mohsenzadeh,\emph{ Non-Linear Trans-Planckian Corrections of Spectra due to the Non-trivial Initial States}, Phys. Lett. B 735 (2014) 261-265.
\bibitem{Tak32} J. P. Gazeau, J. Renaud and M. V. Takook, \emph{Gupta-Bleuler quantization for minimally coupled scalar fields in de Sitter space}, Class. Quant. Grav. 17, 1415 (2000), [gr-qc/9904023].\\
    H. Pejhan, M. V. Takook and M. R. Tanhayi, \emph{Casimir Effect For a Scalar Field via Krein Quantization}, Annals of Physics 341 (2014),[math-ph/1204.6001].\\
          M. Mohsenzadeh, A. Sojasi and E. Yusofi,\emph{ Spectrum of Gravitational Waves in Krein Space Quantization}, Mod. Phys. Lett. A 26, 2697 (2011),[gr-qc/1202.4975];\\
\bibitem{Bun33} T. S. Bunch and P. C. W. Davies, \emph{Quantum field theory in de Sitter space-Renormalization by point-splitting}, Proc. R. Soc. Lond. A  117, 360 (1978).
\bibitem{asl34}   E. D. Stewart, D. H. Lyth, \emph{A more accurate analytic calculation of the spectrum of cosmological perturbations produced during inflation},  Phys. Lett. B (1993) [gr-qc/9302019].

\bibitem{tra36} J. Martin and R. H. Brandenberger, \emph{The TransPlanckian problem of inflationary cosmology}, Phys.Rev. D63 (2001) 123501, [hep-th/0005209].

\bibitem{tra37} H. Collins and R. Holman, \emph{Trans-Planckian enhancements of the primordial non-Gaussianities}, Phys.Rev. D80 (2009) 043524, [arXiv:0905.4925], A. Ashoorioon, A. Kempf, R.B. Mann, \emph{Minimum Length Cutoff in Inflation and Uniqueness of the Action}, Phys.Rev. D71 (2005) 023503, [astro-ph/0410139].

\bibitem{tra38} R. Easther, B. R. Greene, W. H. Kinney, and G. Shiu, \emph{Inflation as a probe of short distance physics}, Phys.Rev. D64 (2001) 103502, [hep-th/0104102].
\bibitem{tra39} R. Brandenberger and P. M. Ho, \emph{Noncommutative space-time, stringy space-time uncertainty principle, and density fluctuations}, Phys.Rev. D66 (2002) 023517, [hep-th/0203119].

\bibitem{tra40} F. Lizzi, G. Mangano, G. Miele, and M. Peloso, \emph{Cosmological perturbations and short distance physics from noncommutative geometry}, JHEP 0206 (2002) 049, [hep-th/0203099].
\bibitem{tra41} R. Easther, B. R. Greene, W. H. Kinney, and G. Shiu,\emph{A Generic estimate of transPlanckian modifications to the primordial power spectrum in inflation}, Phys.Rev. D66 (2002) 023518, [hep-th/0204129].
\bibitem{tra42} U. H. Danielsson, \emph{Inflation, holography, and the choice of vacuum in de Sitter space}, JHEP 0207 (2002) 040, [hep-th/0205227].
\bibitem{tra43} N. Kaloper, M. Kleban, A. E. Lawrence, and S. Shenker, \emph{Signatures of short distance physics in the cosmic microwave background}, Phys.Rev. D66 (2002) 123510, [hep-th/0201158].
\bibitem{tra44} U. H. Danielsson, \emph{A note on inflation and transplanckian physics}, Phys.Rev. D66 (2002) 023511, [hep-th/0203198].
\bibitem{mal45} N. Agarwal, R. Holman, A. J. Tolley and J. Lin,\emph{Effective field theory and non-Gaussianity from general inflationary states}
, JHEP 1305 (2013) 085, [arXiv:1212.1172];\\ J. M. Maldacena, \emph{Non-Gaussian features of primordial fluctuations in single field inflationary models}, JHEP 0305 (2003) 013, [astro-ph/0210603];\\
     L. Senatore, \emph{TASI 2012 Lectures on Inflation}, Published by World Scientific Publishing Co. Pte. Ltd., 2013.
\bibitem{tra45} K. Goldstein and D. A. Lowe, \emph{Initial state effeects on the cosmic microwave background and transPlanckian physics}, Phys.Rev. D67 (2003) 063502, [hep-th/0208167].
\bibitem{Reg46} D.M. Regan, \emph{Measuring CMB non-Gaussianity as a probe of Inflation and Cosmic Strings}, PhD Thesis, DAMTP, Cambridge(2011), [astro-ph/1112.5899].
\bibitem{man46} M. Mohsenzadeh, E. Yusofi and M. R. Tanhayi, Particle creation with excited de Sitter modes, Canadian Journal of Physics, 2015, 93(12): 1466-1469, 10.1139/cjp-2015-0294.
\bibitem{map47}  G. Hinshaw et al. [WMAP Collaboration], Astrophys. J. Suppl. 208, 19 (2013) [astro-ph:1212.5226];
 \\ C. Cheng and Q. Huang,  \emph{Constraint on inflation model from BICEP2 and WMAP 9-year data}, [astro-ph/1404.1230].


\end{thebibliography}
\end{document}